%% file: STA.tex
\documentclass[dvipsnames]{article}

\usepackage[margin=1in]{geometry}

\usepackage{amssymb}
\usepackage{graphicx,amsmath,upgreek,bm}
\usepackage[T1]{fontenc}
\usepackage{colortbl}
\usepackage{wrapfig}
\usepackage[rgb, table]{xcolor}
\usepackage{booktabs}
\usepackage{makecell}
\usepackage[toc,page]{appendix}
\usepackage{multirow}
\usepackage{algorithm}
\usepackage{rotating}
\usepackage{wrapfig}
\usepackage{lineno,hyperref}
\modulolinenumbers[5]
\hypersetup{
   colorlinks=true,
   linkcolor=blue,
   filecolor=magenta,
   urlcolor=blue,
   citecolor=blue,
	}
\usepackage[
  separate-uncertainty = true,
  multi-part-units = repeat
]{siunitx}
\usepackage{caption} 
\usepackage{subcaption}
\usepackage{appendix}    
\usepackage{booktabs} 
\usepackage[para,online,flushleft]{threeparttable}
\usepackage[noend]{algpseudocode}

\usepackage{arydshln}
\makeatletter
  \renewcommand*\env@matrix[1][*\c@MaxMatrixCols c]{%
    \hskip -\arraycolsep
    \let\@ifnextchar\new@ifnextchar
  \array{#1}}
\makeatother
\newcommand\correspondingauthor{\thanks{Corresponding author.}}

\makeatletter
\def\BState{\State\hskip-\ALG@thistlm}
\makeatother
\definecolor{light-gray}{gray}{0.9}

\title{Automated Surface Texture Analysis via Discrete Cosine Transform and Discrete Wavelet Transform}

\author{Melih C. Yesilli\correspondingauthor\\
				Department of Mechanical Engineering\\
				Michigan State University\\
				yesillim@msu.edu
			\and
				Jisheng Chen\\	
				Department of Mechanical Engineering\\
				Michigan State University\\
				chenjish@msu.edu
			\and
				Firas A.~Khasawneh\\
				Department of Mechanical Engineering\\
				Michigan State University\\
				khasawn3@egr.msu.edu
			\and
				Yang Guo\\
				Department of Mechanical Engineering\\
				Michigan State University\\
				yguo@msu.edu
				}
				
\date{}

\begin{document}
\maketitle

\begin{abstract}
Surface roughness and texture are critical to the functional performance of engineering components. The ability to analyze roughness and texture effectively and efficiently is much needed to ensure surface quality in many surface generation processes, such as machining, surface mechanical treatment, etc. 
Discrete Wavelet Transform (DWT) and Discrete Cosine Transform (DCT) are two commonly used signal decomposition tools for surface roughness and texture analysis. 
Both methods require selecting a threshold to decompose a given surface into its three main components: form, waviness, and roughness. 
However, although DWT and DCT are part of the ISO surface finish standards, there exists no systematic guidance on how to compute these thresholds, and they are often manually selected on case by case basis. 
This makes utilizing these methods for studying surfaces dependent on the user's judgment and limits their automation potential. 
Therefore, we present two automatic threshold selection algorithms based on information theory and signal energy. 
We use machine learning to validate the success of our algorithms both using simulated surfaces as well as digital microscopy images of machined surfaces. 
Specifically, we  generate feature vectors for each surface area or profile and apply supervised classification. 
Comparing our results with the heuristic threshold selection approach shows good agreement with mean accuracies as high as 95\%. 
We also compare our results with Gaussian filtering (GF), and show that while GF results for areas can yield slightly higher accuracies, our results outperform GF for surface profiles.  
We further show that our automatic threshold selection has significant advantages in terms of computational time as evidenced by decreasing the number of mode computations by an order of magnitude compared to the heuristic thresholding for DCT. 
\end{abstract}


\input{Intro}
\input{simulation}
\input{methods}

\input{results_and_conclusion}

%
\bibliography{STA}
\end{document}

%% file: Intro.tex
\section{Introduction}
\label{sec:Intro}
Enhancements in measurement technology have opened the door for applying surface texture analysis to various applications such as medical imaging \cite{Wu1992,Taha_2015,Martisius2018}, construction materials \cite{Tong2018}, remote sensing \cite{Rignota1990,Makarenko2016} and tribology \cite{Myshkin2003,Suh2003}.
Some of the challenges in surface texture analysis include the data size and the computational effort of some of the current methods. 
Specifically, as the resolution of the surface images increases, their size also increases, which makes the data processing cumbersome and computationally expensive.  
Therefore, there is a need for increasing automation and decreasing the computational complexity of algorithms for surface texture analysis. 
Another challenge is to extract appropriate descriptors for a surface in an automated way.
The most common approach used in the literature is to decompose data into form, waviness, and roughness components. 
This approach is applied to both profiles \cite{Raja2002,Raja1979,Dobrzanski2010,Lee1998} and surfaces~\cite{Lecompte_2010,Goic2016,Goic2011}.
The form component includes low-frequency content in the surface or profiles, waviness involves the mid-range frequencies, while the high frequencies are collected in the roughness component. 
In general, the form and the waviness of a surface scan or profiles are obtained, then they are subtracted from the original data to obtain the roughness component. 

Gaussian filter is one of the widely adopted signal processing tools for surface roughness analysis \cite{ISO16610, Raja2002, Gurau2004, Hendarto2006,Janecki2011}.
It is used to smooth the surface profile measurement to obtain an approximation of the raw surface profile. 
The mean line is then subtracted from the measurement to obtain a roughness profile.
Raja et al.~used a Gaussian filter to obtain an approximation to surface profiles, and they compared this approximation with the ones obtained from the 2RC filter, one of the earliest filters used for surface metrology \cite{Raja2002}. 
Hendarto et al. focus on the roughness analysis of wood surface using Gaussian filter \cite{Hendarto2006}.
However, the main drawback for Gaussian filtering approach is the boundary distortion where the mean of the end parts of a surface profile cannot be used ~\cite{Raja2002}. Raja et al. suggested that the end parts of the mean line should be ignored for evaluation~\cite{Raja2002}, while this is not feasible for profiles with shorter lengths. Therefore, Janecki proposed a solution that extrapolates both end of profiles with polynomial functions to eliminate the edge effect~\cite{Janecki2011}.
Another approach used in the analysis of profiles of engineering surfaces is Fast Fourier Transform (FFT). 
Raja and Radhakrishnan used FFT to obtain the surface roughness by removing the lower frequency components of form and waviness. 
Dong et al.~provide an extensive understanding of two-dimensional FFT (2D-FFT) analysis on engineering surfaces \cite{Dong1995}.
Peng et al.~used 2D-FFT to identify the type of the wear particles on surfaces using angular spectrum values which are obtained by converting Cartesian coordinates into polar form \cite{Peng1997}. 
Empirical Mode Decomposition (EMD), one of the most commonly adopted signal decomposition tools, is another approach used for the analysis of engineering surfaces. 
Several versions of EMD are proposed to analyze surfaces such as Bidimensional EMD (BEMD) \cite{Nunes2003}, Image EMD (IEMD)\cite{LINDERHED2009}, Bidimensional Multivariate EMD (BMEMD) \cite{Xia2019}. However, the computation of EMD in 2D is slow compared to other approaches.

Discrete Cosine Transform (DCT) is another widely used approach for decomposing a surface scan into its form, waviness, and roughness components~\cite{Lin2008, Lecompte_2010, Chandankhede2011, Goic2016}. 
Lecompte et al.~developed an approach to identify the form and the contribution of classical defects such as positioning error and tool deflection \cite{Lecompte_2010}. They used only a certain percentage of the DCT coefficients to obtain a filtered surface. However, when we have a large number of images, each image may require the usage of a different percentage of the DCT coefficients to generate the form. 
In general, DCT requires selecting two threshold values for delineating the three different components of the surface. 

Discrete Wavelet Transform (DWT) is another approach used extensively for surface texture analysis~\cite{Chen1995,Liu1996,Raja2002,Goic2016,Fu2003,Josso2002, Morala2012,Chang2005,Wang2017}. Chen et al.~introduced DWT for surface profiles \cite{Chen1995}. Liu et al.~obtained a threshold that isolates the form of the surface by computing all possible approximations that can be obtained using the coefficients at each level. Another example of this approach is seen in Ref.~\cite{Raja2002, Fu2003} where the separation of the three components of a profile is performed using multi-resolution analysis approximations. 
The common procedure is to apply the DWT at a certain level and obtain the approximation and detail coefficients, and then use the approximation coefficients for the reconstruction of the form component \cite{Chen1999}.
The detail coefficients are then used to reconstruct waviness and roughness. 
Nevertheless, there is a need for a guideline on how to automatically choose the threshold that separates the mid-frequency content from the higher ones in the DWT approach. 
In addition, selection of the mother wavelet function can also affect the resulting components. Stkepien et al. used autocorrelation, cross-correlation, and entropy-based test to evaluate the performance of different wavelet functions used in surface texture analysis~\cite{Stkepien2015}.

To our knowledge, there is no approach for automatically separating the form, waviness, and roughness components for DCT and DWT, and the current practice is to manually select them using the user's experience and judgment call. 
Therefore, we propose an automatic, data-driven approach for identifying the needed thresholds for DCT and DWT. 
For DWT, we utilize the energy of the reconstructed signals to separate the waviness and roughness from each other, while for DCT we leverage the surface entropy to define the form and waviness components. Roughness is then found by subtracting the filtered surface from the original one. 

In addition to our contributions to the automatic threshold selection in DCT and DWT, we identify the machining processes through which surface samples are generated. Most studies in the literature are focused on small patch processes with few samples where human interpretation is heavily used to identify and compute profile or surface roughness. 
In contrast, we validate our approach for a large data set obtained from simulated surfaces, as well as experimental surface scans. 
We obtain the roughness components of surfaces and profiles using our automatic threshold algorithms and we extract the 1D and 2D features introduced in the ISO standards~\cite{ISO25178, ISO21920}. 
We then utilize machine learning to assess the accuracy of the automatic thresholds. 
Specifically, we use the obtained features in supervised classification algorithms to classify surfaces that are labeled with respect to the generating surface parameter for the simulated surfaces, and with respect to the generating machining process for the experimental data.  
We use support vector machine (SVM), logistic regression (LR), random forest (RF), and gradient boosting (GB) algorithms for classification and employ hyperparameter tuning using the grid search approach. 

This article is organized as follows. 
Section~\ref{sec:simulation} provides the details for synthetic surface generation, experimental data collection, and data preprocessing. 
Section~\ref{sec:Methods} introduces our new automatic threshold selection algorithms and explains how to use them for feature extraction.
Section~\ref{sec:Results} provides classification results obtained using the proposed algorithms and compares them to the results obtained by heuristic threshold selection. 
We provide our concluding remarks in Sec.~\ref{sec:conclusion}. 

%% file: simulation.tex
\section{Data Collection}
Our data includes both synthetic surfaces (Section~\ref{sec:simulation}), and digital scans of machined surfaces (Section~\ref{sec:Exp_data_collection}. The following subsections provide more information on each data type.  
\subsection{Synthetic surface generation}
\label{sec:simulation}
We used the model described in Ref.~\cite{M_ser_2017} to generate synthetic surfaces. 
The roughness level of the surfaces is controlled by varying the Hurst exponent $H$, which takes parameter values between 0 (rough surface) and 1 (smooth surface).
We chose 201 different $H$ values in this range, and for each $H$ value we generated a different surface. 
We then categorized these surfaces with respect to their roughness level. 
For instance, the first 67 surfaces were labeled smooth, while the last 67 surfaces were considered rough.
The surfaces in between these two extremes were considered somewhat rough. 
The generated surfaces were then used to obtain both areal and profile features. 
Profiles of the generated surfaces were obtained by taking cross-sections along the generated surface's $x$ and $y$ directions, and they were assigned the same labels as the original surface. 
Depending on the type of signal processing tool used, we obtained roughness surfaces or roughness surface profiles, and then we extracted the corresponding features needed for the classification algorithms. 

\subsection{Experimental Data Collection and Preprocessing}
\label{sec:Exp_data_collection}
A standard S-22 Microfinish Comparator is used for physical surface texture data collection. To better recognize the surface texture, 9 sample surfaces with clearly observed rough texture on the comparator are selected (see Tab.~\ref{tab:selected_surfaces}).

\begin{table}[h]
\centering
\caption{Selected surfaces under various machining conditions}
\label{tab:selected_surfaces}
\begin{tabular}{c|ccc}
\multicolumn{1}{c|}{Machining Type}&\multicolumn{3}{c}{Roughness height (micrometers(microinches))}\\
\hline
M - milling & 3.175 (125) & 6.35(250) & 12.7(500)\\
P - profiled &3.175 (125) & 6.35(250) & 12.7(500)\\
ST - shaped or turned & 3.175 (125) & 6.35(250) &  12.7(500)\\
\hline
\end{tabular}
\end{table}

\begin{figure}
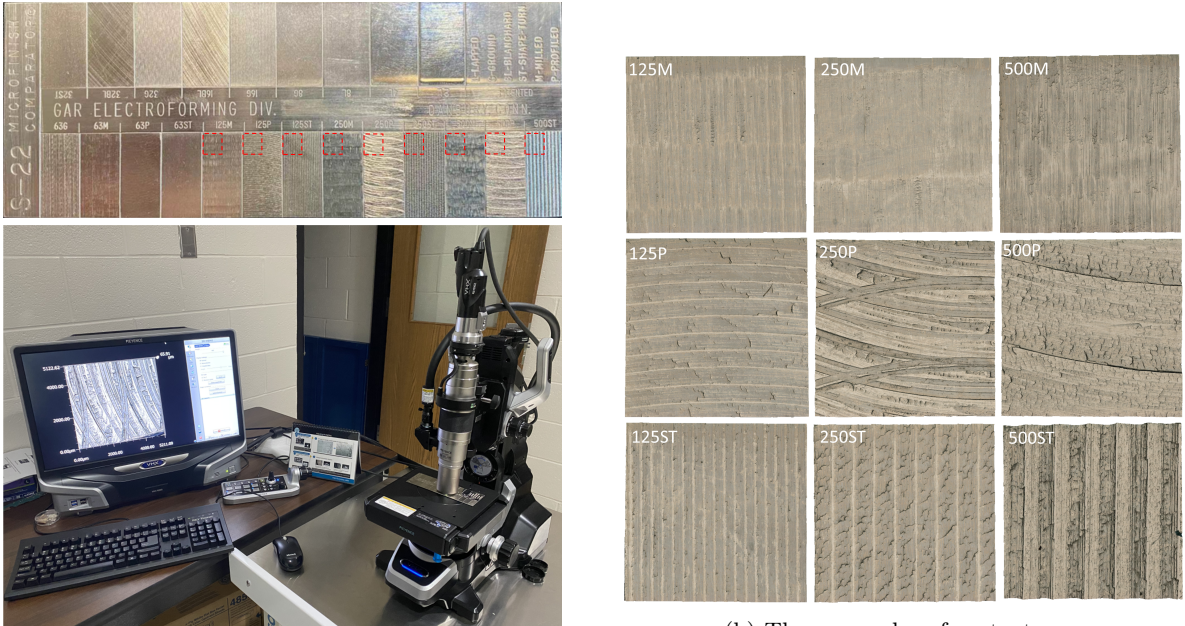

\centering
\begin{subfigure}{.5\textwidth}
  \centering
  \includegraphics[width=.9\linewidth]{keyance-device}
  \caption{Scanned portions of the sample and the digital microscope.}
  \label{fig:microscope}
\end{subfigure}%
\begin{subfigure}{.5\textwidth}
  \centering
  \includegraphics[width=.9\linewidth]{scanned_surface}
  \caption{The scanned surface textures.}
  \label{fig:scanned_surfaces}
\end{subfigure}
\caption{The microscope used for experimental data collection, and the sample surfaces.}
\end{figure}

The scanned area is 5 mm $\times$ 5 mm, and it was consistently located at the upper left corner for each measured sample surface. 
Specifically, the microfinish comparator was placed on a free-angle XYZ motorized observation system (VHX-S650E), and the surface textures were measured using a Keyence digital microscope  (VHX6000),  as shown in Fig.~\ref{fig:microscope}. A real zoom lens (Keyence VH-Z500R, RZ x500-x5000) is used to capture the surface texture and $\times500$ magnification is used to achieve sufficient spatial resolution (0.42 $\mu$m). The stitching technique (11 $\times$ 11 scans in the horizontal and vertical directions) is performed to enlarge the observation view so that the whole selected area can be captured under this magnification. 
The spatial sampling rate of the images was approximately 2.4 samples per $\mu$m. 
Figure~\ref{fig:scanned_surfaces} shows the resulting scanned surfaces. 

\subsection{Data Preprocessing}
\begin{wrapfigure}{r}{0.4\textwidth}
    \centering
    \vspace{-10pt}
    \includegraphics[width=0.4\textwidth,height=1\textheight,keepaspectratio]{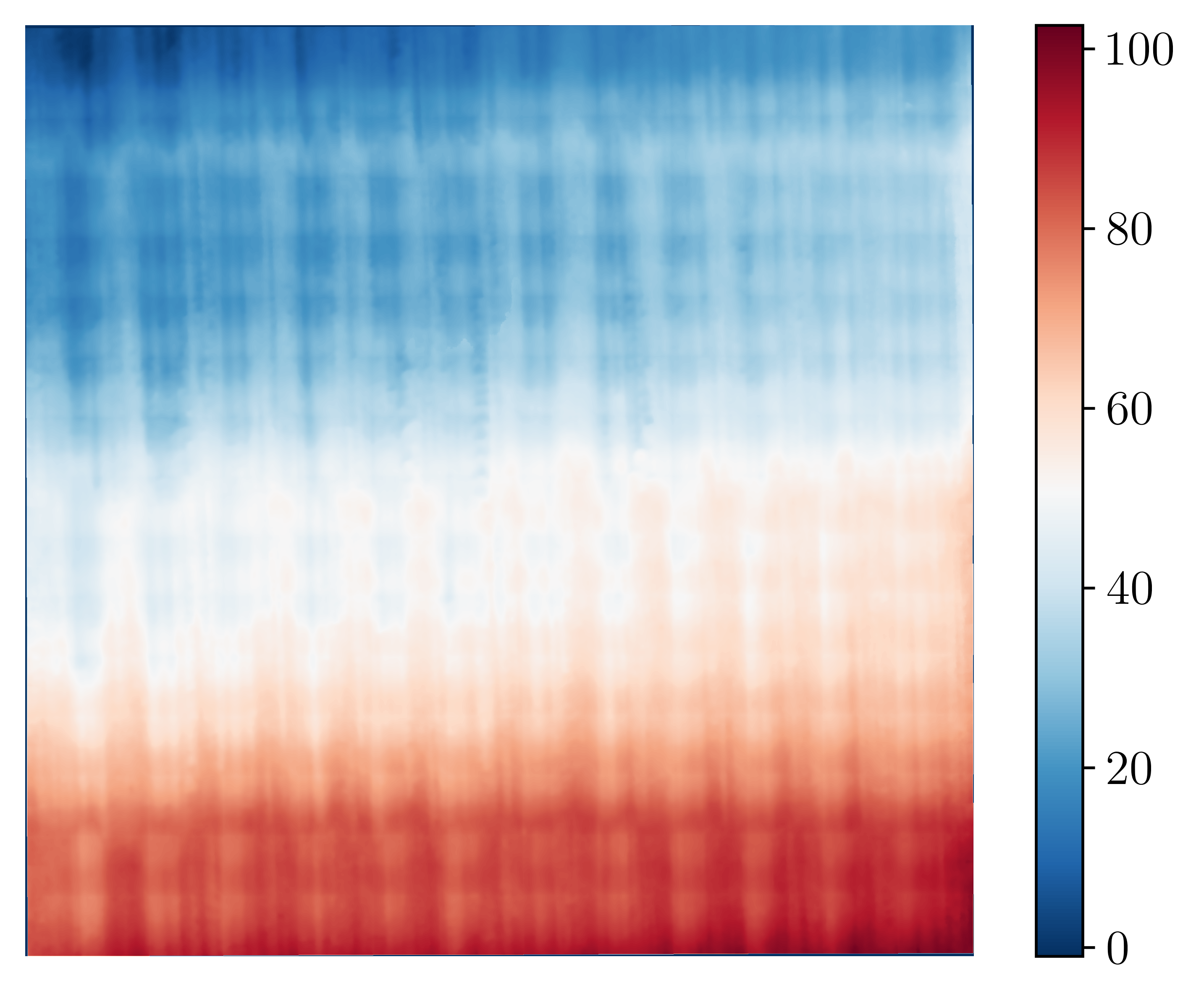}
    \caption{Scan sample of 125M.}
    \label{fig:125M_scan}
\end{wrapfigure}
The resulting raw surface scans include different numbers of pixels with lower gray-level intensity values on the edges of the image as shown in Fig.~\ref{fig:125M_scan}. 
These pixels are tedious to isolate manually, so we copped the images and adaptively removed these pixels using the following algorithm. 
First, we found the remainder of image pixel values in each direction when they are divided by 1000. 
The halves of the remainders are used as the number of pixels to remove from each edge. 
This procedure was successful in significantly reducing the number of pixels with lower gray-level intensity values at the boundaries, and the resulting images had similar sizes. 

The other challenge was the large dimension of the microscope surface scans which can exceed 10000 pixels in each direction of the image thus elevating computational expenses. 
Consequently, we split each surface scan into 25 sub-images each with a dimension of 2400$\times$2400 pixels.  

\subsubsection{Image Subsampling}
The resulting subimages still presented computational challenges for some signal processing tools such as DCT where the maximum number of modes is equal to the total number of pixels. 
Therefore, we subsampled the images to further reduce the number of samples in the subimages when using 2D signal processing tools for surface classification. 
Several approaches are available for image scaling/resampling in the literature. 
These also include some signal processing approaches to upscale or downscale an image. 
One of the simple and widely used subsampling methods is to replace a block of pixels by their average values, and that is the approach we used in this study. 
After testing different sampling factors such as 0.1, 0.2, and 0.5, we adopted a sampling factor of 0.1 for the experimental data set. This means the size each block is $10\times 10$ pixels.


%% file: methods.tex
\section{Methods}
\label{sec:Methods}

\subsection{Discrete Wavelet Transform}
\label{sec:DWT}
\begin{figure}[!h]
\centering
\includegraphics[width=0.9\textwidth,height=1\textheight,keepaspectratio]{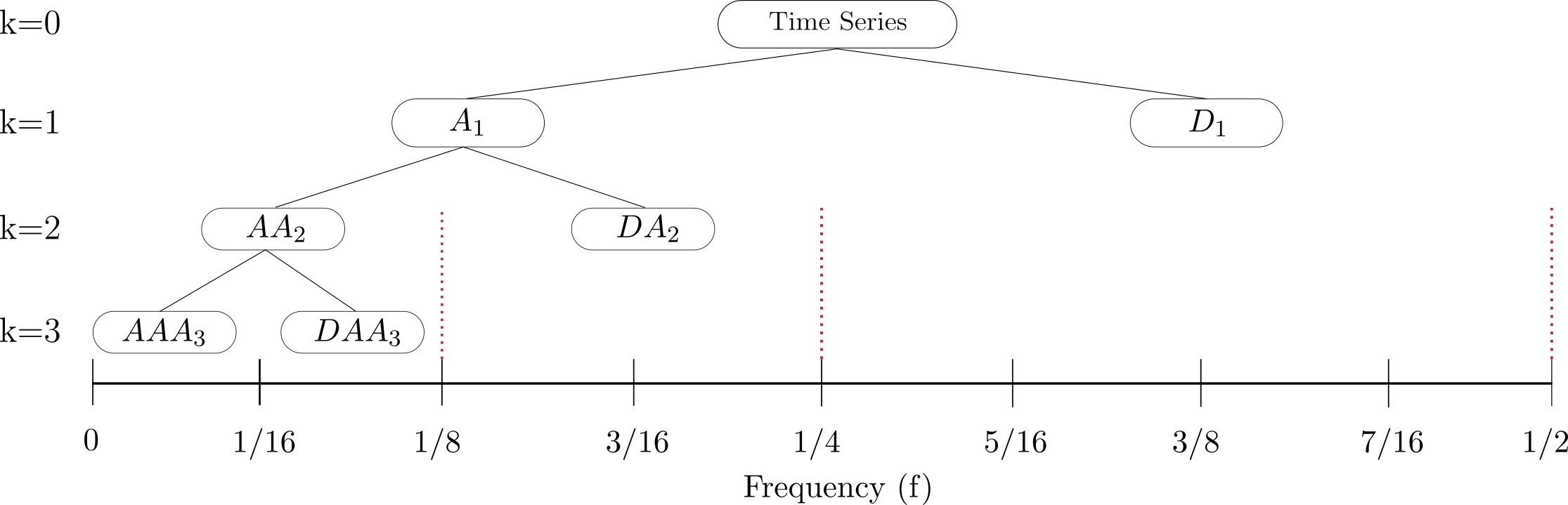}
\caption{DWT tree for the first three level of the transform.}
\label{fig:DWT_tree}
\end{figure}

Discrete Wavelet Transform (DWT) is one of the widely adopted signal processing tools \cite{Gao2010, Yesilli2020a, Olkkonen2011, Prabhakar2002, Gurley1999}.
While a signal's frequency spectrum can only be represented over the entire time domain with Fourier Transform, Wavelet Transform can decompose the signal into components with different time and frequency resolutions \cite{Newland2012}.
In DWT, the time series is passed through low pass and high pass filters to obtain approximation and detail coefficients. 
These filters are obtained from a filter bank derived from scaling and translating a wavelet function \cite{Gao2010}. 
The diagram for Discrete Wavelet Transform is provided in Fig.~\ref{fig:DWT_tree} where the level of the transform is denoted as $k$. 
As the level increases, only approximation coefficients are passed through the filters to obtain new approximation and detail coefficients.
For example, the approximation coefficient of the first level transform $A_{1}$ is passed through the filters, and we obtain new approximation and detail coefficients as $AA_{2}$ and $DA_{2}$. As we increase the level of the transform, we increase the frequency resolution of decomposition.
As seen from Fig.~\ref{fig:DWT_tree}, each component corresponds to a distinct frequency range.

In this implementation, we utilize DWT to analyze simulated and experimental surface profiles. 
1D DWT is applied to surface profiles with biorthogonal wavelet functions (BIOR4.4) recommended by the ISO standards~\cite{ISO16610-29}. The specified wavelet functions are symmetric and surface profiles can be reconstructed without any loss~\cite{ISO16610-29}.
The level of the transform is chosen based on the maximum allowable limit defined by the number of samples in the profile and the type of the wavelet function.
Approximation coefficients at the maximum level of the transform are used to obtain the form of the profile \cite{Chen1999}. 
The waviness and roughness profiles of the surface are reconstructed using the detail coefficients of the transform.
For instance, $DAA_{3}, DA_{2}$ and $D_{1}$ can be used to reconstruct profiles for waviness and roughness when 3 levels of DWT are employed (see Fig.~\ref{fig:DWT_tree}).
Profile reconstructed from $AAA_{3}$ represents the form of the profile.
The separation between waviness and roughness is done heuristically in the literature, and to our knowledge, there is not any guide on how to select a threshold for detail coefficients.
Therefore, we describe an approach that leverages signal energy to automatically select that threshold. 

\begin{figure}[!h]
\centering
\includegraphics[width=1\textwidth,height=1\textheight,keepaspectratio]{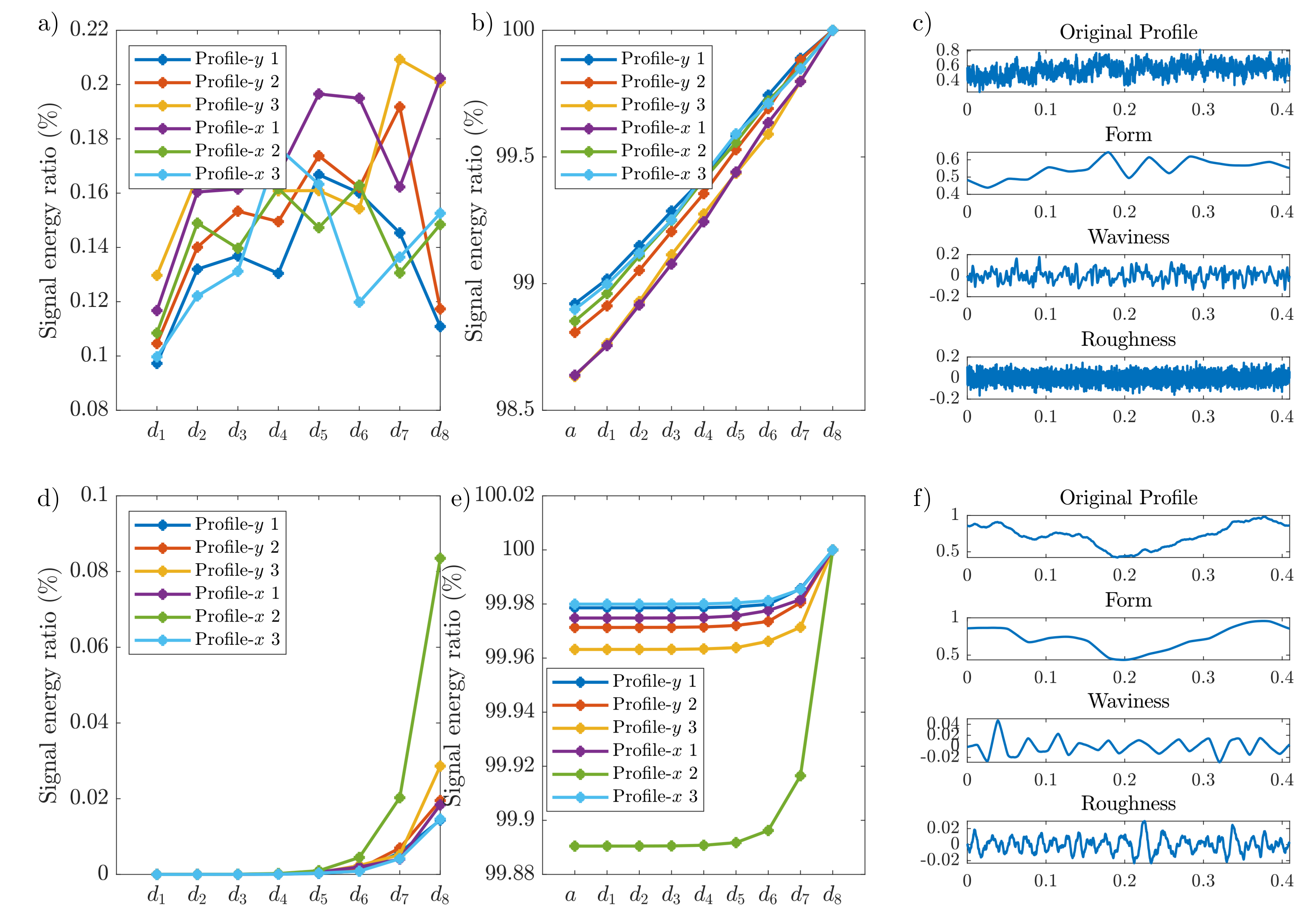}
\caption{The plots on the first row belong to roughest surface ($H=0$) simulation, while the plots in the second row belong to smoothest surface ($H=1$) a) Energy ratios of detail coefficients ($d_{i}$: detail coefficients of level $i$). b) Cumulative energy ratios including approximation coefficients ($a$: approximation coefficients of maximum level). c) The resulting three main components after applying the automatic threshold selection for for profile $x-1$.}
\label{fig:signal_energy_DWT}
\end{figure}

The energy of a discrete signal is defined as $E = \sum_{n=-\infty}^{\infty} \mid x[n]\mid^{2}$.
After applying DWT, we reconstruct profiles using the detail coefficients at all levels and compute their signal energy.
Figure~\ref{fig:signal_energy_DWT} provides the energy ratio of each detail coefficient and the cumulative energy ratios in addition to the decomposition obtained with automatic threshold selection.
The first row of the plots is obtained with the roughest surface, and the ones in the second row belong to the smoothest surface. 
Both surfaces have a size of $4096\times 4096$ pixels.
We take three cross-sections in each direction of the surfaces represented with Profile-$x$ (or $y$) $i$ in the figure.

\begin{figure}[!h]
\centering
\includegraphics[width=1\textwidth,height=1\textheight,keepaspectratio]{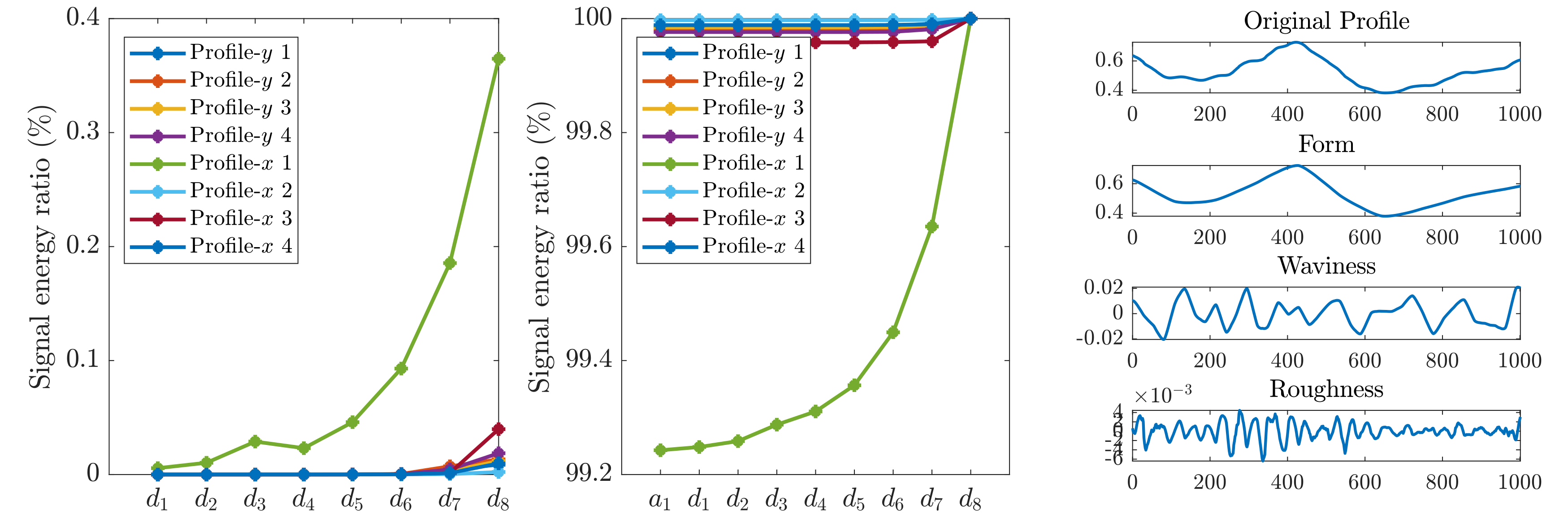}
\caption{(left) Energy ratios of detail coefficients ($d_{i}$: detail coefficients of level $i$). (middle) Cumulative energy ratios including approximation coefficients ($a$: approximation coefficients of maximum level). (right) The resulting three main components after applying the automatic threshold selection for profile $x-1$.}
\label{fig:signal_energy_DWT_experimental}
\end{figure}

The cumulative energy ratio plot is given in Figs.~\ref{fig:signal_energy_DWT}b and~\ref{fig:signal_energy_DWT}e for the smoothest and the roughest surface of the simulation.
It is seen that most of the energy is accumulated in approximation coefficients.
Since we are interested in separating waviness and roughness, we only take into account the energy ratios of detail coefficients.
In Fig.~\ref{fig:signal_energy_DWT}a, it is hard to notice the significant increase in signal energy as we increase the level of the transform.
Therefore, we use the differences between consecutive energy ratios.
For example, the biggest difference between consecutive energy ratios for Profile-$y$ 1 in Fig.~\ref{fig:signal_energy_DWT}a is between levels 4 and 5. 
Our algorithm uses detail coefficients of the first four levels to reconstruct roughness.
The rest of the detail coefficients are reconstructed and summed up to obtain a waviness profile.
The resulting surface components are shown in Fig.~\ref{fig:signal_energy_DWT}c.
When we decrease the roughness of the surface, the threshold is easier to capture since we see a dramatic increase in signal energy between levels 7 and 8 (see Fig.~\ref{fig:signal_energy_DWT}d) for most of the profiles.
In this case, roughness is equal to the summation of reconstructed signals from detail coefficients of the first seven levels, and the $8$th level coefficients are used to generate the waviness profile.
In Fig.~\ref{fig:signal_energy_DWT_experimental}, we provide energy plots of the reconstructed signals. Cumulative energy plots show that the signals obtained using approximation coefficients have the highest energy. It is evident that the form profile, which is obtained from approximation coefficients, has the largest amplitudes in the decomposition. The significant increase in energy occurs between levels 7 and 8. Therefore, for this example, the threshold is selected at level 7. This approach is applied to all profiles extracted from the sub-images in the experimental data, and the features are computed accordingly. 

\subsection{Discrete Cosine Transform}
\label{sec:DCT}
Discrete Cosine Transform (DCT) decomposes a signal into cosine functions with different frequencies. 
It is similar to Fourier Transform except that it uses only real coefficients. 
There are several types of DCT, and in this implementation, we use Type II DCT. 
The definition of 2D transform is given as~\cite{Strang1999}
\begin{equation}
B_{pq} = \alpha_{p}\alpha_{q}\sum_{m=0}^{M-1}\sum_{n=0}^{N-1}I_{mn}cos\Bigg(\frac{\pi(2m+1)p}{2M}\Bigg)cos\Bigg(\frac{\pi(2n+1)q}{2N}\Bigg),
\label{eq:forward_DCT_2D}
\end{equation}
where 
\[ \alpha_{p} = \begin{cases} 
      \frac{1}{\sqrt{M}} & p=0 \\
      \sqrt{\frac{2}{M}} & 1 \leq p \leq M-1 \\
   \end{cases}  \quad \text{and} \quad
	\alpha_{q} = \begin{cases} 
      \frac{1}{\sqrt{N}} & q=0 \\
      \sqrt{\frac{2}{N}} & 1 \leq q \leq N-1 \\
   \end{cases}.  
\]
$M$ and $N$ are the numbers of elements in each direction of the image, and $I_{mn}$ represents the image.
The inverse transform is provided as
\begin{equation}
I_{ij} =  \sum_{p=0}^{M-1}\sum_{q=0}^{N-1}\alpha_{p}\alpha_{q}B_{pq}cos\Bigg(\frac{\pi(2i+1)p}{2M}\Bigg)cos\Bigg(\frac{\pi(2j+1)q}{2N}\Bigg).
\label{eq:inverse_DCT_2D}
\end{equation}
If we insert Eq.~\eqref{eq:forward_DCT_2D} into Eq.~\eqref{eq:inverse_DCT_2D}, the resulting expression is
\begin{equation}
\resizebox{1\textwidth}{!}{$
A_{ij} =  \sum_{p=0}^{M-1}\sum_{q=0}^{N-1}\alpha_{p}\alpha_{q}\Bigg(\alpha_{p}\alpha_{q}\sum_{m=0}^{M-1}\sum_{n=0}^{N-1}A_{mn}cos\Bigg(\frac{\pi(2m+1)p}{2M}\Bigg)cos\Bigg(\frac{\pi(2n+1)q}{2N}\Bigg)\Bigg)
cos\Bigg(\frac{\pi(2i+1)p}{2M}\Bigg)cos\Bigg(\frac{\pi(2j+1)q}{2N}\Bigg).
$}
\label{eq:inverse_DCT_2D_modified}
\end{equation}
\setlength{\columnsep}{5pt}%

\begin{figure}
\centering
\begin{subfigure}{.5\textwidth}
  \centering
  \includegraphics[width=1\linewidth]{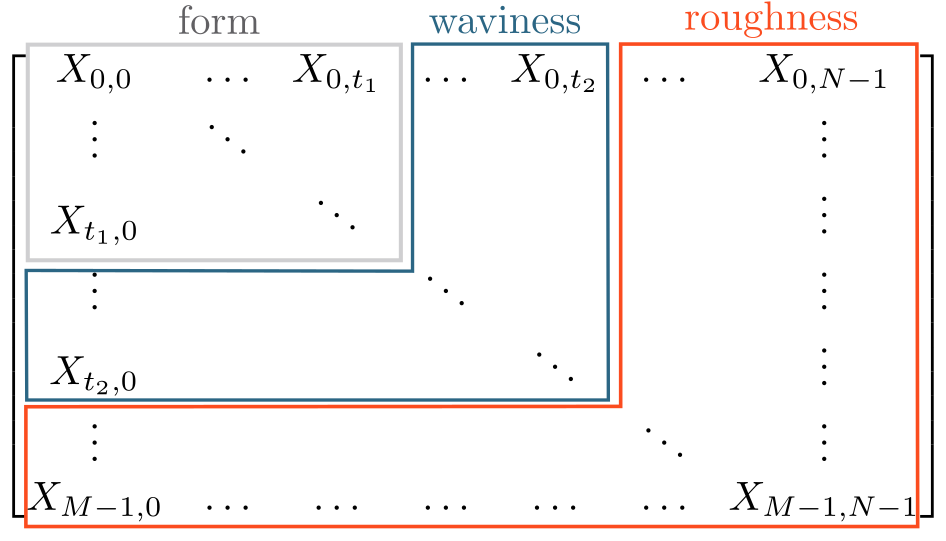}
  \caption{DCT modes in matrix format}
  \label{fig:modes_in_matrix}
\end{subfigure}%
\begin{subfigure}{.5\textwidth}
  \centering
  \includegraphics[width=1\linewidth]{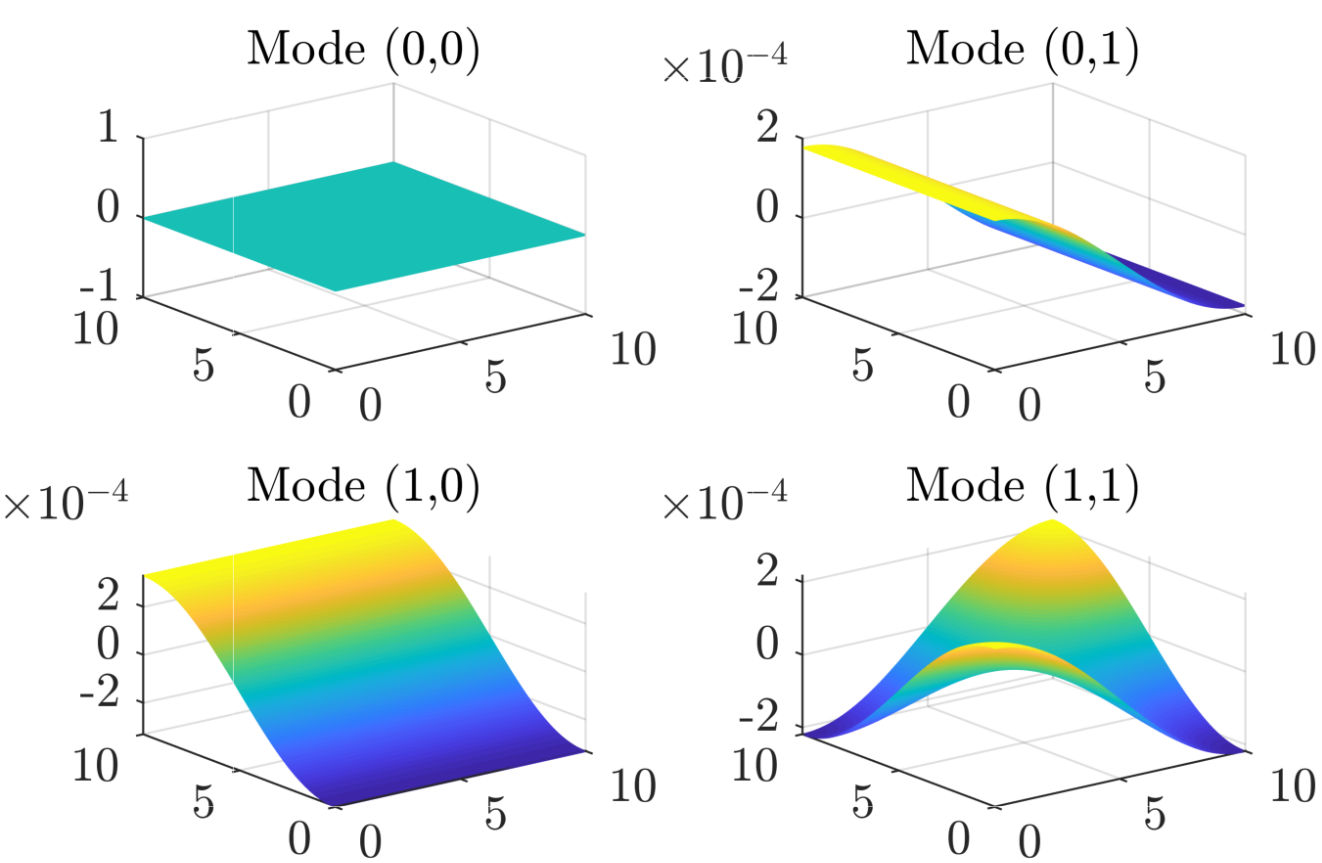}
  \caption{The first four modes of a synthetic surface.}
  \label{fig:example_modes}
\end{subfigure}
\caption{Matrix format of the DCT modes and example modes.}
\end{figure}

Basis functions (modes of a given image) are defined with indices $p$ and $q$. 
Summing all the modes recovers the original image as shown in Eq.~\eqref{eq:inverse_DCT_2D_modified}. 
In Fig.~\ref{fig:modes_in_matrix}, the modes are represented by $X_{i,j}$. For instance, $X_{0,0}$ represents the first mode of a given surface. 
The first four modes of a synthetic surface are provided in Fig.~\ref{fig:example_modes}. 
Form, waviness, and roughness components are obtained by summing the modes inside the boxes shown with grey, blue, and red colors, respectively.
Figure~\ref{fig:modes_in_matrix} shows that we need to select two thresholds $t_{1}$ and $t_{2}$ to separate the three components. 
Since we are interested in finding the surface roughness component, we neglect $t_{1}$ and treat form and waviness as one combined component. 
We then introduce a new approach based on image entropy to generate an automatic threshold selection for DCT, as described in the next section. 

\subsubsection{Threshold Selection Using Image Entropy}
Information entropy is known as Shannon's entropy \cite{Shannon1948} and it is defined as
\begin{equation*}
H(X) = \sum_{i=1}^{n} p_{i}log(1/p_{i}),
\end{equation*}
where $X$ is a discrete random variable with probability distribution $p$.  
An analogous expression is obtained for computing the image entropy \cite{Sabuncu2006} according to
\begin{equation}
H(I) = \sum_{i}(h_{I}(i)/N)log(N/h_{I}(i)),
\end{equation} 
where $h_{I}(i)$ is the counts in histogram of the grayscale image $I$, and $N$ is the number of bins the histogram.

\begin{wrapfigure}{r}{0.5\textwidth}
 \vspace{-25pt}
  \begin{center}
    \includegraphics[width=0.5\textwidth]{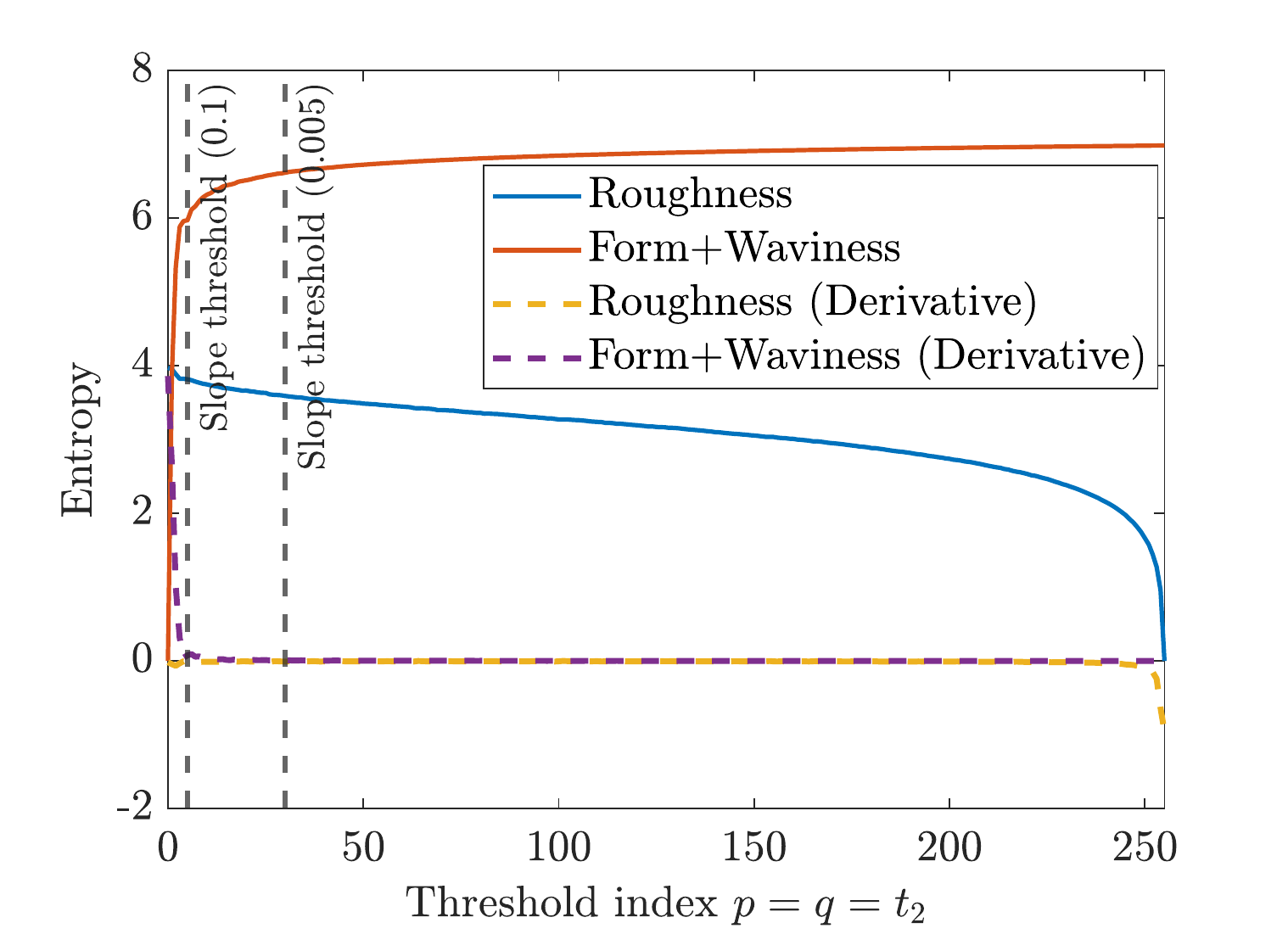}
  \end{center}
  \vspace{-20pt}
  \caption{Entropy of the roughness and waviness+form surface for varying threshold index.}
  \label{fig:entrpy}
  \vspace{-15pt}
\end{wrapfigure}
As mentioned earlier, we compute two components for a surface and these are the roughness component and the surface that includes waviness and form components. 
We start with $t_{2}=0$. 
The first mode $X_{0,0}$ combines the form and waviness components of the surface, while the rest of the modes represent the roughness component. 
Combining form and waviness also gives us the advantage of avoiding additional mode computation for the surface. 
We can obtain the roughness component by simply subtracting the first component (form+waviness) from the original image.
After obtaining both components, we compute their image entropies. 
Then, we increase the threshold index $t_{2}$ by one. 
Now, the first four modes shown in Fig.~\ref{fig:example_modes} represent the first component (form+waviness).
The roughness component is obtained by subtracting the first component from the original image, and image entropy for both components is computed.
This procedure is iterated by increasing the threshold index $t_{2}$.
We repeated this process for all threshold indices from 0 to 256 for the roughest surface in the data set, and we obtained entropy curves for both components as shown in Fig.~\ref{fig:entrpy}.

\begin{figure}[h]
\vspace{-10pt}
\centering
\includegraphics[width=1\textwidth,height=1\textheight,keepaspectratio]{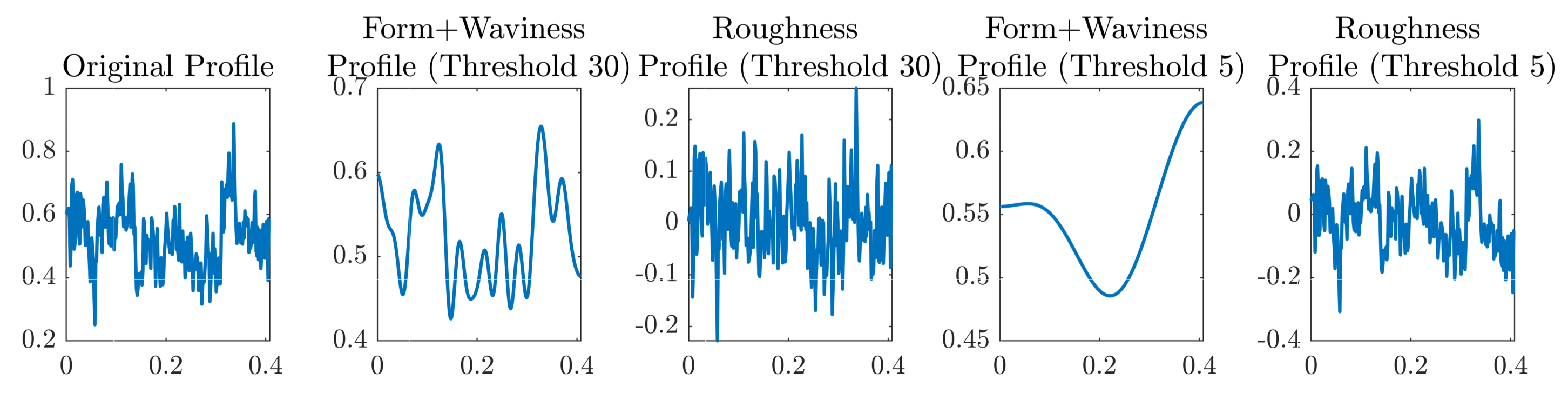}
\caption{Reconstructed surface profiles obtained using thresholds from the entropy analysis shown in Fig.~\ref{fig:entrpy}.}
\label{fig:recon_ent_thld}
\end{figure}

Figure~\ref{fig:entrpy} shows that the entropy of the first component (form+waviness) is increasing dramatically for a small change of $t_{2}$. 
After a certain value of the threshold, its rate of increase slows down as seen from its derivative curve. 
Therefore, we set a threshold value for the slope of the entropy curve of the first form+waviness component. 
We tested the two threshold values 0.1 and 0.005 (see Fig.~\ref{fig:entrpy}, grey vertical lines).
When the slope of the curve is under these slope thresholds, their corresponding $t_{2}$ values are taken as thresholds to separate the roughness component. 
For these two slope thresholds 0.1 and 0.005, $t_{2}$ is found as 5 and 30, respectively.
Then, we obtained two components of the surface using these threshold values as shown in Fig.~\ref{fig:recon_ent_thld}.
The figure shows only surface profiles to clearly illustrate the difference between the components. 
It is seen that threshold 30 is more reasonable to use since the form+waviness component obtained with $t_{2}=5$ does not provide a good approximation to the surface. 
Therefore, we decided to use slope threshold 0.005 for all surfaces in the data set.

An automatic threshold selection algorithm is defined such that it can terminate the loop when the slope of the entropy curve of the form+waviness component is below 0.005.
For instance, we only compute the first $30^{2}$ modes of the surface whose profiles are given in Fig.~\ref{fig:recon_ent_thld} instead of computing all $256^{2}$ modes.
Therefore, we avoid unnecessary and expensive mode computations.
This algorithm is applied to both synthetic and experimental data sets to obtain roughness components that can be used to compute the 2D features needed for applying machine learning. 

\subsection{Gaussian Filter}
Gaussian smoothing is another approach for analyzing surfaces (or profiles) using the convolution of the surface (or profile) with a kernel. One dimensional implementation of Gaussian filtering is based on a kernel defined as~\cite{ISO16610}
\begin{equation*}
K(x) = \frac{1}{\alpha \lambda_{c}}\text{exp}\Bigg(-\pi\Big(\frac{x}{\alpha\lambda_{c}}\Big)\Bigg),
\end{equation*}
where $\alpha$ is $\sqrt{2/\pi}$, and $\lambda_{c}$ is the long wavelet cutoff. This cutoff is decided based on the procedure provided in Ref.~\cite{ASMEB46}. The roughness value $Ra$ of a profile is calculated. Then, we select a cutoff from Table 3-3.20.2-1 in Ref.~\cite{ASMEB46}. A new $Ra$ is then calculated using the chosen cutoff value. 
If the new $Ra$ is outside of corresponding $Ra$ range of the cutoff value, a new cutoff is selected according to the newest $Ra$.
However, the selected cutoff value may be larger than the measurement length of the profile. In this case, the algorithm keeps the current cutoff value instead of using a new one. 
We set the cutoff using this procedure and obtain a smoother surface profile. This filtered profile is subtracted from the original one, and the resulting profile constitutes the surface's roughness profile. 

For the implementation of Gaussian filtering on images, we use a kernel defined as
\begin{equation}
    K(x,y) = \frac{1}{2\pi \sigma^{2}}\text{exp}\Bigg(\frac{-x^{2}-y^{2}}{2\sigma^{2}}\Bigg).
\end{equation}
The convolution operation is defined as 
\begin{equation}
    I[m,n] = \sum_{u=-W}^{W}\sum_{v=-W}^{W}K[u,v]f[m-u,n-v],
\end{equation}
where $I$ represents the smoothed surface, $f$ is the original surface. The kernel size for Gaussian smoothing is defined as $2\times W+1$ and denoted as $K$. The standard deviation $\sigma$ is selected using $\sigma = K/6$. We used a kernel size of 21 in all computations. The resulting smooth surfaces are subtracted from the original ones to obtain the roughness. Then, we computed the 2D features from this component to generate feature matrices for supervised classification algorithms. We used this approach to compare the results obtained from the algorithms that we defined in Sec.~\ref{sec:DWT} and~\ref{sec:DCT}. 

\subsection{Feature extraction from surfaces}
\label{sec:feature_extraction}
This section describes the feature extraction procedure from surfaces. 
We use the roughness component of each surface or profile.
Depending on the type of data we use, we compute the profile or areal parameters given in Ref.~\cite{ISO21920, ISO25178}.
While working with surface area measurements, we compute the height and hybrid parameters provided in Secs.~4.2 and 4.4 of \cite{ISO25178} for roughness components of the surfaces. 
For surface profiles, we computed height, spatial and hybrid parameters given in Secs.~4.1-4.2 of \cite{ISO21920}.
The list of features and their definitions are provided in Tab.~\ref{tab:features}. Definitions for all the features are given in continuous form. 
The discrete form of these equations can be found in Ref.~\cite{ISO25178} and~\cite{ISO21920}.
Then, we feed these feature matrices to supervised classification algorithms, namely, Support Vector Machine (SVM), Logistic Regression (LR), Random Forest (RF), and Gradient Boosting (GB).
In this study, we used the default parameters of the classifiers. 
 
\begin{table}[!h]
\centering
\caption{The features used in the classification of surfaces and surface profiles. $f(x)$ and $f(x,y)$ represent surface profile and surface, respectively.}
\label{tab:features}
\resizebox{1\textwidth}{!}{
\begin{tabular}{c|c}
\multicolumn{1}{c|}{1D Features}&\multicolumn{1}{c}{2D Features}\\
\hline
$Rq$ = $\sqrt{\frac{1}{l_{m}}\int_{l_{m}}f(x){2}dx}$, \; $Rsk$ = $\frac{1}{R_{q}^{3}}\frac{1}{l_{m}}\int_{l_{m}}f(x)^{3}dx$ &$Sq$ = $\sqrt{\frac{1}{A}\iint_{\tilde{A}}f^{2}(x,y)dxdy}$\\

$Rku$ = $\frac{1}{R_{q}^{4}}\frac{1}{l_{m}}\int_{l_{m}}f(x)^{4}dx$, \; $Rt$ = $\text{max}(f(x))+|\text{min}(f(x))|$ & $Ssk$ = $ \sqrt{\frac{1}{AS_{q}^{3}}\iint_{\tilde{A}}f^{3}(x,y)dxdy}$ \\

$Ra$ = $\frac{1}{l_{m}}\int_{l_{m}}|f(x)|dx $& $S_{ku}$ $= \frac{1}{AS_{q}^{4}}\iint_{\tilde{A}}f^{4}(x,y)dxdy$\\

$Ral$ = $\min\limits_{t_{x}\in R} t_{x},\:\text{where}\: R = \{t_{x}: \text{ACF}(t_{x})<s\} $& $Sp$ $= \max(f(x,y))$\\

$Rsw$ = $\frac{2\pi}{arg\,\max\limits_{p}|F(p)|}$, \; $Rdt$ = $\max\limits_{x\in R}|\frac{dz(x)}{dx}|$& $Sv$ = $|\min(f(x,y))|$ \\

$Rdq$ $= \sqrt{\frac{1}{l_{m}}\int_{l_{m}}\Bigg(\frac{df(x)}{dx}\Bigg)^{2}dx}$& $Sz$ = $Sp+Sv$\\

$Rda$ = $\frac{1}{l_{m}}\int_{l_{m}}|\frac{df(x)}{dx}|dx$ & $Sa$ $= \frac{1}{A}\iint_{\tilde{A}}|f(x,y)|dxdy$\\

$Rdl$ = $\int_{l_{m}}\Bigg(\sqrt{1+\Bigg(\frac{df(x)}{dx}\Bigg)^{2}}\Bigg)dx$&
$Sdq$ $= \sqrt{\frac{1}{A}\iint_{\tilde{A}}\Bigg[\Bigg(\frac{\partial f(x,y)}{\partial x}\Bigg)^{2}+\Bigg(\frac{\partial f(x,y)}{\partial y}\Bigg)^{2}\Bigg]dxdy}$
\\

$Rdr$ = $\frac{1}{l_{m}}\int_{l_{m}}\Bigg(\sqrt{1+\Bigg(\frac{df(x)}{dx}\Bigg)^{2}}-1\Bigg)dx$&
$Sdr$ = $\frac{1}{A}\iint_{\tilde{A}}\sqrt{\Bigg[1+\Bigg(\frac{\partial f(x,y)}{\partial x}\Bigg)^{2}+\Bigg(\frac{\partial f(x,y)}{\partial y}\Bigg)^{2}\Bigg]}-1\Bigg)dxdy$\\
\end{tabular}}
\end{table}

%% file: results_and_conclusion.tex
\section{Results and Discussion}
\label{sec:Results}

This section presents the results obtained with automatic threshold selection algorithms introduced in Sec.~\ref{sec:Methods}, and compares the results to the ones obtained from the Gaussian filter.
We apply both algorithms to the synthetic and experimental data sets described in Sec.~\ref{sec:simulation}, and we classify the surfaces and their profiles.
We have three labels for synthetic surfaces representing the surface's roughness level. 
For experimental data, we have samples coming from three main machining operations each of which corresponds to three different surface roughness ranges. 
Therefore, we have nine labels for the experimental data set. We also perform three-class classification by only taking into account the type of the machining operation. 
In addition to using the automatic threshold selection algorithms, we heuristically decompose surfaces into form, waviness, and roughness. 
Specifically, we manually inspect the decomposition of a few surfaces or profiles and make a decision on the threshold values mentioned in Sec.~\ref{sec:DWT} and~\ref{sec:DWT}.
These thresholds are then fixed and used for the whole data set.
The resulting roughness surfaces or profiles are used to extract features as explained in Sec.~\ref{sec:feature_extraction}, and supervised classification is performed.

\begin{figure}[!h]
\centering
\includegraphics[width=1\textwidth,height=1\textheight,keepaspectratio]{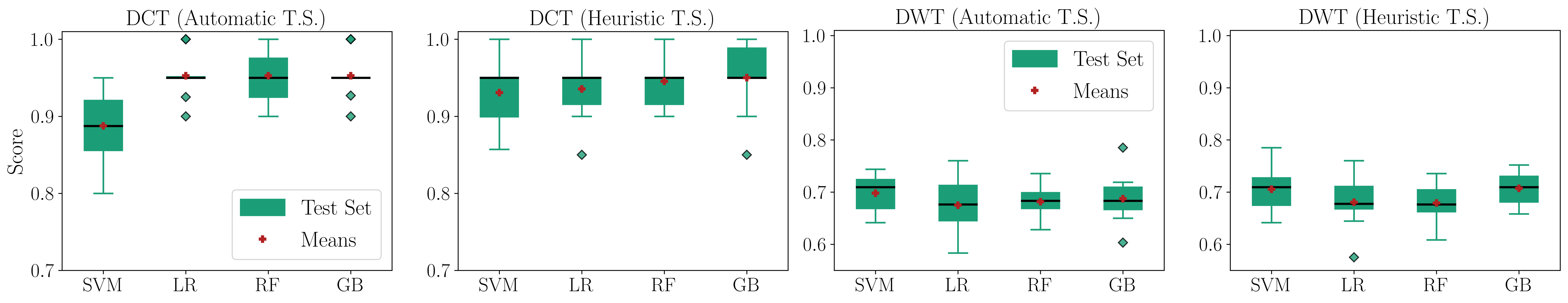}
\caption{Classification accuracies obtained with automatic and heuristic threshold selection for DCT and DWT using synthetic data set. Three class classification performed in this case.}
\label{fig:DCT_DWT_Comparison}
\end{figure}

Fig.~\ref{fig:DCT_DWT_Comparison} provides the figures for the classification results obtained from DCT and DWT using automatic and heuristic threshold selection. It is seen that there is no significant difference between the classification accuracies of automatic threshold selection and heuristic threshold selection.
We see that the automatic threshold algorithm helps to decrease the deviation in the results obtained with LR and GB classifiers for DCT.
One should note that heuristic threshold selection requires manual inspection of several surfaces to decide the threshold value. 
Depending on the number of surfaces inspected, the manual process can add a significant amount of time for identifying the roughness level. 
In contrast, automatic threshold selection algorithms for both DWT and DCT do not require manual inspection since we only need to choose a slope threshold as explained in Sec.~\ref{sec:DCT}.

\begin{figure}[!h]
\centering
\includegraphics[width=1\textwidth,height=1\textheight,keepaspectratio]{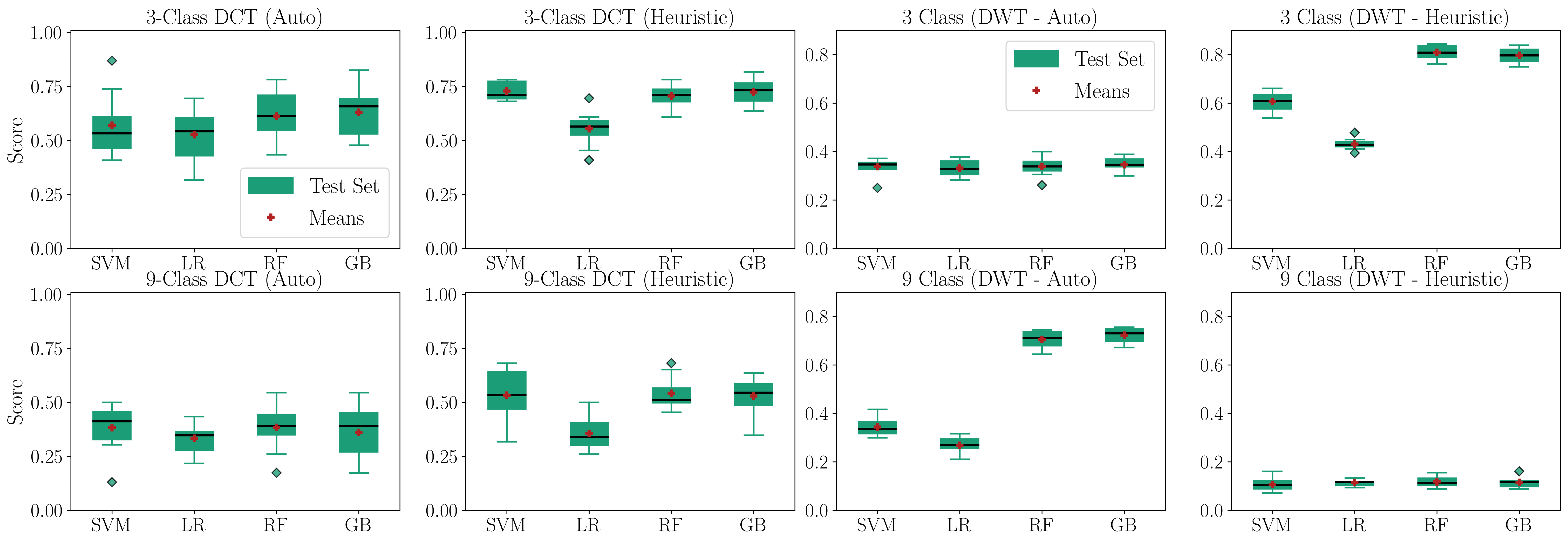}
\caption{Experimental data results for DWT (profile classification) and DCT (Surface classification).}
\label{fig:Exp_Results_DWT_DCT}
\end{figure}

Experimental data results for DWT and DCT are provided in Fig.~\ref{fig:Exp_Results_DWT_DCT}. It includes the results obtained with the automatic threshold selection algorithms and the heuristic approach where we select the components by manual inspection.
The first row of the figure provides plots for 3-class classification where the labels are milling (M), profiling (P), and shaped or turned (ST). The second row contains the results obtained with the nine-class classification where each machining operation has three different roughness values.
It is seen that DCT automatic threshold selection provides similar accuracies compared to the heuristic approach.
\begin{wrapfigure}{r}{0.40\textwidth}
 \vspace{-20pt}
  \begin{center}
    \includegraphics[width=0.40\textwidth,height=1\textheight,keepaspectratio]{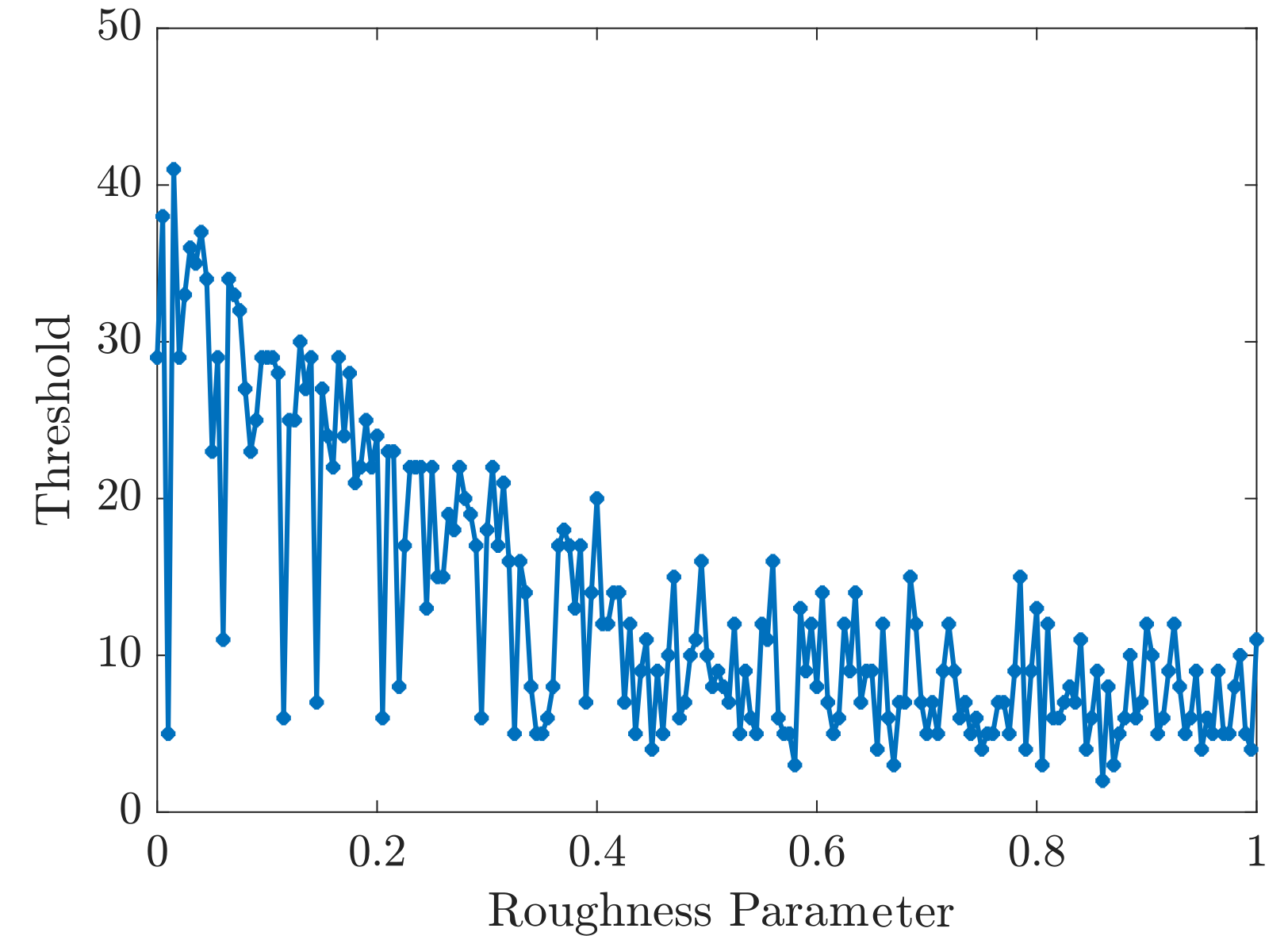}
  \end{center}
  \vspace{-10pt}
  \caption{The thresholds selected by our algorithm for the synthetic data set. Roughness parameter $0$ represents the roughest surface, while the smoothest surface has roughness parameter of $1$.}
  \label{fig:Thresholds}
  \vspace{-15pt}
\end{wrapfigure}
We also performed nine-class classification by assigning distinct labels to the main surfaces shown in Fig.~\ref{fig:scanned_surfaces}.
Fig.~\ref{fig:Exp_Results_DWT_DCT} shows that nine classification provides better classification scores.
When we compare the results of the heuristic selection of thresholds and the automated approach, DCT results are slightly lower when we compared them to the heuristic approach.
However, nine-class classification with automatic threshold selection for DWT outperforms the heuristic threshold selection (see Fig.~\ref{fig:Exp_Results_DWT_DCT}).
Nevertheless, three-class classification with the heuristic approach of DWT provides better accuracies.
This can be explained by the fact that the selected threshold is suitable for a large portion of the data set. 
Since the threshold selection is highly dependent on the person who performs the manual inspection, the heuristic approach results may not be consistent.  

\begin{figure}[!hb]
\centering
\includegraphics[width=1\textwidth,height=1\textheight,keepaspectratio]{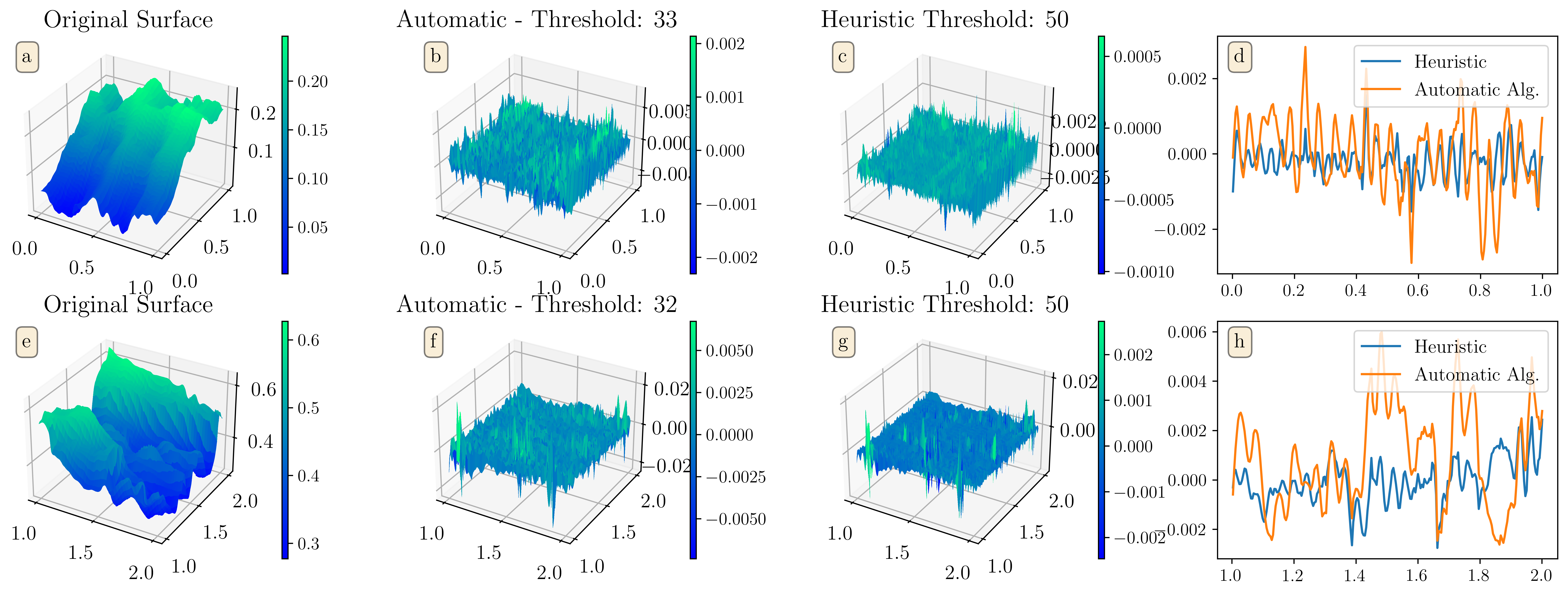}
\caption{Roughness surfaces and roughness profiles obtained from three surfaces in experimental data set. (a-d) Milling, (e-h) Profiled.}
\label{fig:Heuristic_Automatic_Comparison}
\end{figure}

\begin{figure}[!h]
\centering
\includegraphics[width=0.5\textwidth,height=1\textheight,keepaspectratio]{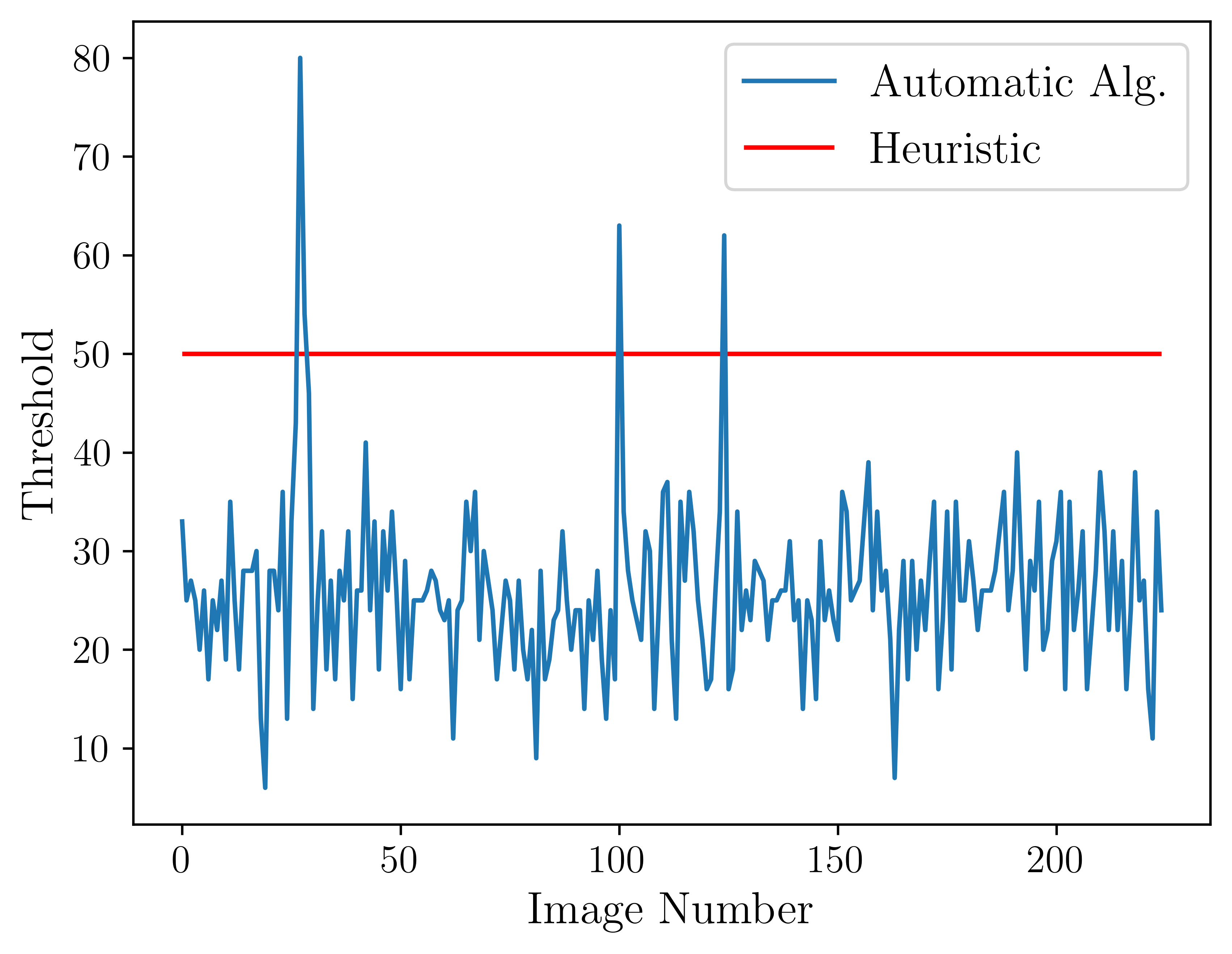}
\caption{Threshold values selected from automatic threshold selection algorithm and the constant heuristic threshold.}
\label{fig:thresholds_Experimental}
\end{figure}

Manual threshold selection for DWT and DCT is performed using visual inspection. 
One may choose a large value for the threshold depending on the size of the given surfaces.
In our case, we selected the thresholds of DCT as 50 for both synthetic and experimental data sets, while thresholds of DWT were chosen as 2 and 4 for synthetic and experimental data sets, respectively.
We provide the roughness components of three surfaces obtained from the heuristic approach and the proposed automatic threshold selection algorithms in Fig.~\ref{fig:Heuristic_Automatic_Comparison}. 
It is seen that roughness profiles obtained from the heuristic approach have smaller amplitudes compared to the ones obtained from the automatic threshold selection approach. 
This is because of the fact higher thresholds remove more modes from the main surface and this leads to less number of modes for the roughness component.
In addition, the selected thresholds are less than the heuristic thresholds value which we keep constant for all surfaces in the experimental data set. In Fig.~\ref{fig:thresholds_Experimental}, we also provide the selected threshold values from the automatic selection algorithm and the constant heuristic threshold.

Our threshold selection algorithms remove the manual inspection and make the decomposition fully automatic. For DCT, our approach provides a significant reduction in computational time. Since the threshold is selected as 50 for the heuristic DCT approach, we would need to compute $50^{2}$ surface modes for each surface in the data set. However, the automatic threshold selection algorithm picks different values of the threshold depending on the surface. Figure~\ref{fig:Thresholds} provides the selected threshold values for the synthetic surfaces for varying roughness parameters.
It is seen that the maximum selected threshold is nearly 40. That means that the algorithm computes $40^{2}$ modes at maximum for the corresponding surface. Compared to the number of mode computations performed with the heuristic threshold, we save the time needed to compute 900 modes. The time reduction is even more significant with smoother surfaces with a higher value of roughness parameter $H$. 
In this case, we needed to compute $10^{2}$ modes instead of $50^{2}$.
Thus this shows that automatic threshold selection avoids redundant mode computations and dramatically decreases the computational time in comparison to the heuristic threshold selection.

\begin{figure}[!h]
\centering
\includegraphics[width=0.6\textwidth,height=1\textheight,keepaspectratio]{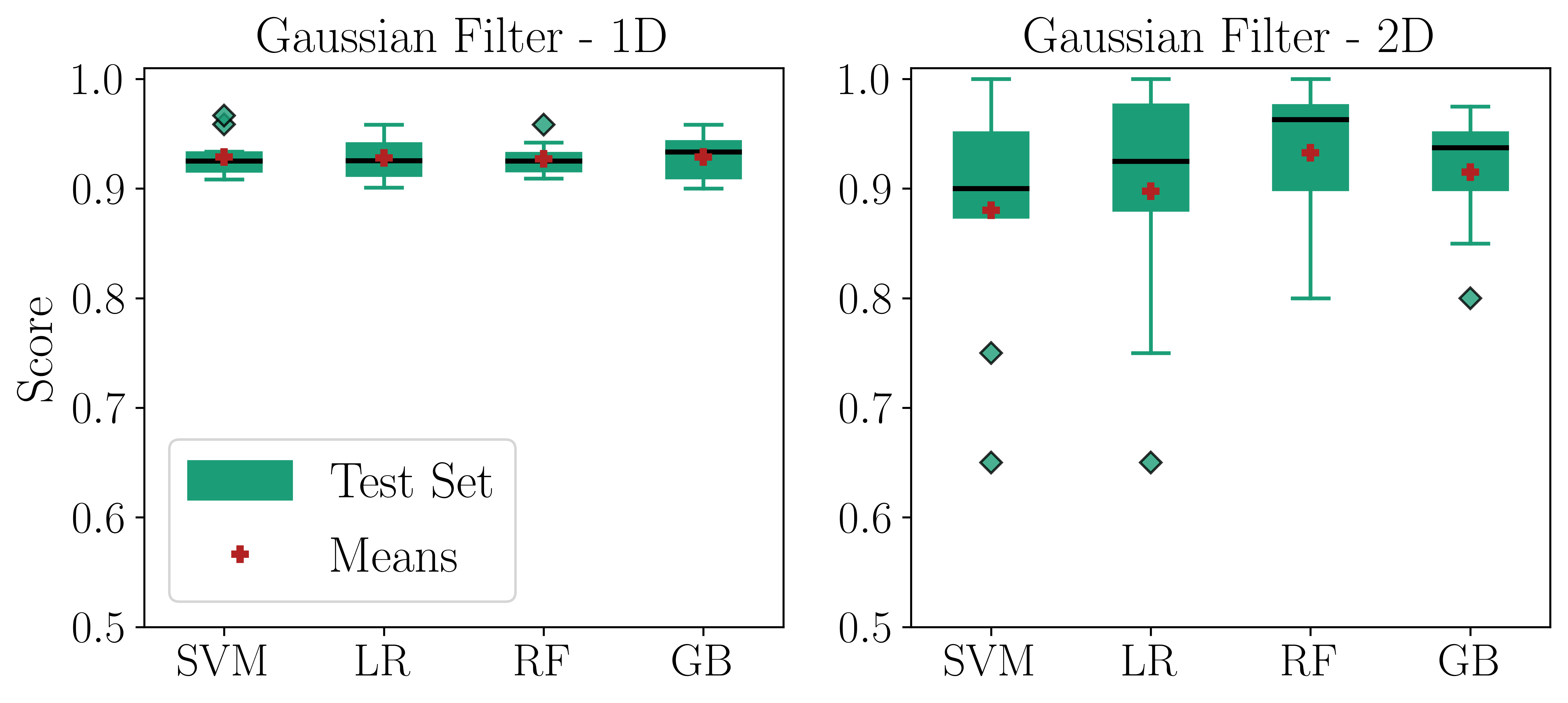}
\caption{Gaussian filtering results obtained from profile and surface classification for synthetic data set.}
\label{fig:GF_1D_2D_Results_Synthetic}
\end{figure}
\begin{figure}[!h]
\centering
\includegraphics[width=0.8\textwidth,height=1\textheight,keepaspectratio]{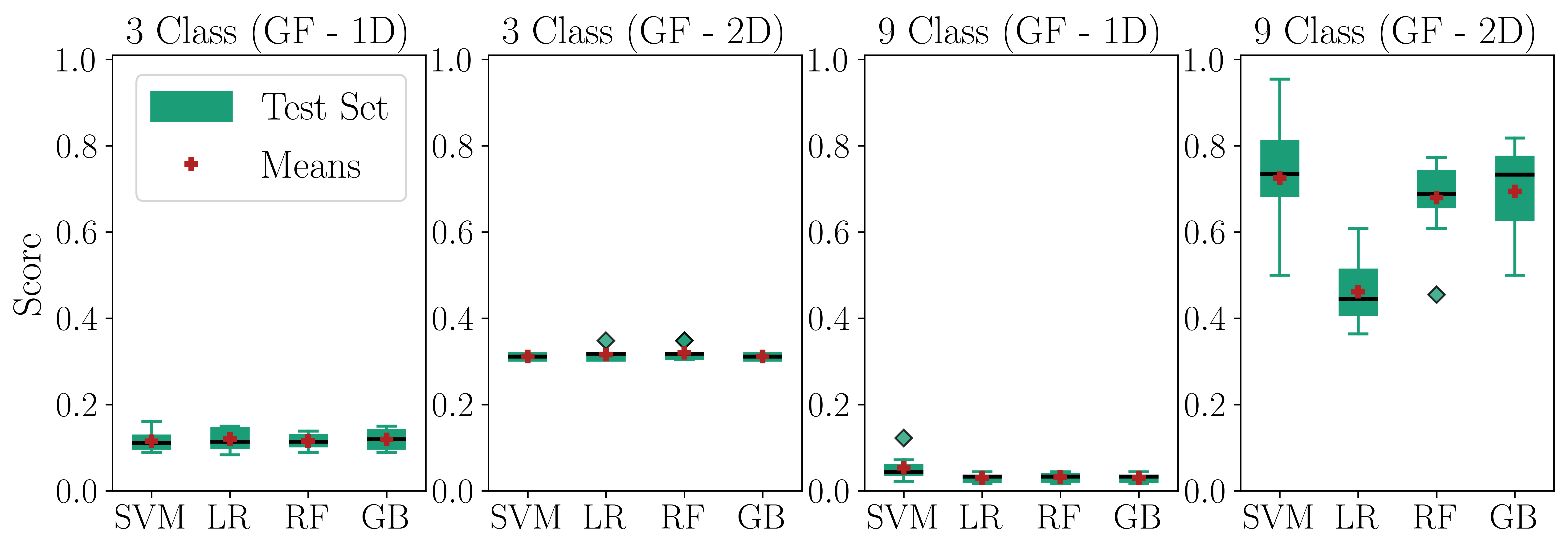}
\caption{Three and nine class classification results for surface profiles (1D) and the surfaces (2D) in experimental data using Gaussian Filter.}
\label{fig:GF_1D_2D_Results}
\end{figure}

We compare our experimental results for DWT and DCT to the ones obtained from Gaussian Filtering, which is another tool widely adopted in digital image and signal processing. 
Figure~\ref{fig:GF_1D_2D_Results_Synthetic} provides classification scores for the Gaussian Filter applied to surface profiles (1D) and surfaces (2D) from the synthetic data set.
Figure~\ref{fig:GF_1D_2D_Results_Synthetic} shows that the highest classification accuracy is obtained with Gaussian Filtering. However, the results for experimental data do not show the same trend. 
Figure~\ref{fig:GF_1D_2D_Results} shows that three-class classification does not perform well. In addition, the nine-class classification for profiles does not perform well either. This can be explained by the fact that the selected number of profiles from the surfaces is not enough to extract texture information. For profiles, we do not apply subsampling, and we extract 8 profiles from the images whose sizes are $2400\times2400$. Therefore, while more profiles may increase the score, exploring that direction is out of the scope of this work. 
Our approaches for DWT provide better accuracies for both profile classification in three and nine-class classification than the ones obtained from Gaussian Filtering (see Fig.~\ref{fig:Exp_Results_DWT_DCT} and Fig.~\ref{fig:GF_1D_2D_Results}).
Nine-class classification for surface classification provides mean accuracies around 70\%, which is higher than the results of DCT shown in Fig.~\ref{fig:Exp_Results_DWT_DCT}.
However, the Gaussian filter when applied in 2D is parameter-dependent, and we used the same parameter (kernel size) for the whole data set in this study. 
We used the same wavelet function for all surface scans, so our algorithm for DWT is parameter independent, while the algorithm for DCT only takes the slope threshold from the user. This threshold value can be set to a value near zero and can be used for the whole data set. However, the threshold selection is still data-driven and it depends on the entropy curve of the given surface (see Fig.~\ref{fig:entrpy}).

\section{Conclusion}
\label{sec:conclusion}
We proposed two automatic threshold selection algorithms for two widely used methods, namely DCT and DWT, to identify the appropriate roughness components in synthetic and experimental data sets. 
In contrast to the traditional way of decomposing engineering surfaces, our algorithms are data-driven and they automatically adapt to a given surface to provide the needed threshold values. 
The algorithm for DWT does not require any input parameter selection from the user, while the one for DCT only requires the user to give a slope threshold for the derivative of the image entropy. 
However, this threshold value is generally close to zero and is easy to set, thus not requiring a high level of expertise. 

The classification accuracies obtained from both our automatic selection algorithms for DCT and DWT either exceed or match the accuracies obtained from the heuristic threshold selection. 
This shows that our algorithms are capable of autonomously decomposing the surfaces to extract the appropriate descriptor of the surface roughness. 
We also eliminate human error which may happen during the manual inspection process in heuristic threshold selection. When we compare our results to the ones obtained from Gaussian smoothing, it is seen that our algorithms provide better classification scores in profile classification and they are comparable to their Gaussian smoothing counterparts.  
In addition, our DCT algorithm is capable of eliminating the redundant mode computations for a given surface. 
This considerably reduces the computational time as evidenced by an order of magnitude improvement in DCT mode computations for a single surface. 
In addition, the mode computation for DCT can be parallelized using High-Performance Computing (HPC) tools, and the computational time needed for DCT can further be significantly reduced. 
We hypothesize that the combination of our algorithms with HPC tools will enable our approach to be used in real-time surface characterization applications. Future work includes the investigation of the effect of mother wavelet function selection, and integration of HPC for real-time surface texture analysis.